\documentclass[english]{neuron-SI}

\usepackage{amsmath} 
\usepackage{amssymb}  
\usepackage{color}

\title{The Energy Landscape Underpinning Module Dynamics in the Human Brain Connectome}
\author{Arian Ashourvan \textsuperscript{1,2} \& Shi Gu \textsuperscript{1,3} \& Marcelo G. Mattar  \textsuperscript{1,4} \& Jean M. Vettel  \textsuperscript{1,2,6} \& Danielle S. Bassett {1,4,7}}

\begin{document}
\maketitle
\begin{affiliations}
\item Department of Bioengineering, University of Pennsylvania, Philadelphia, PA, 19104 USA
\item U.S. Army Research Laboratory, Aberdeen Proving Ground, MD 21005 USA
\item Applied Mathematics and Computational Science Graduate Program, University of Pennsylvania, Philadelphia, PA, 19104 USA
\item Department of Psychology, University of Pennsylvania, Philadelphia, PA, 19104 USA
\item Department of Electrical \& Systems Engineering, University of Pennsylvania, Philadelphia, PA, 19104 USA
\item Department of Psychological \& Brain Sciences, University of California, Santa Barbara,CA, 93106 USA
\item To whom correspondence should be addressed: dsb@seas.upenn.edu
\end{affiliations}

\begin{frontpage}
  \item[Running title:] The Energy Landscape Underpinning Module Dynamics in the Human Brain Connectome
  \item[Corresponding Author:] ~\\
Danielle S. Bassett\\
Department of Bioengineering\\
University of Pennsylvania\\
210 S. 33rd Street\\
240 Skirkanich Hall\\
Philadelphia, PA 19104-6321\\
Ph. 215-746-1754\\
Fax 215-573-2071\
\item [Author contributions:]  A.A., J.M.V., and D.S.B developed the project. M.G.M. acquired the data. A.A. analyzed the data. S.G. contributed computational tools and expertise. A.A., J.M.V., and D.S.B. wrote the paper.
\item[keywords:]{energy landscape $|$ maximum entropy model $|$ community structure $|$ modularity $|$ functional brain network $|$ graph theory }

\end{frontpage}

\begin{abstract}

Human brain dynamics can be profitably viewed through the lens of statistical mechanics, where neurophysiological activity evolves around and between local attractors representing preferred mental states. Many physically-inspired models of these dynamics define the state of the brain based on instantaneous measurements of regional activity. Yet, recent work in the emerging field of \emph{network neuroscience} has provided initial evidence that the human brain might also be well-characterized by time-varying states composed of locally coherent activity or functional modules. Here we study this network-based notion of brain state to understand how functional modules dynamically interact with one another to perform cognitive functions. Specifically, we estimate the functional relationships between regions of interest (ROIs) by fitting a pair-wise maximum entropy model to each ROI's pattern of allegiance to functional modules. Local minima in this model represent attractor states characterized by specific patterns of modular structure. Hierarchical clustering of these local minima highlight three classes of ROIs with similar patterns of allegiance to community states. Visual, attention, sensorimotor, and subcortical ROIs tend to form a single functional community (Class-I). The remaining ROIs tend to form a putative executive control community (Class-II) or a putative default mode and salience community (Class-III). We simulate the brain's dynamic transitions between these community states using a Markov Chain Monte Carlo random walk (MCMC). We observe that simulated transition probabilities between basins strongly resemble empirically observed transitions between community allegiance states in resting state fMRI data. The accuracy of our predictions is strongest for the transition probabilities of Class-I ROIs (primary sensorimotor, attention, subcortical), followed by Class-III ROIs (default mode), but to a lesser extent Class-II ROIs (executive). These results highlight the transient fluctuations characteristic of cognitive control systems. More broadly, these results collectively offer a view of the brain as a dynamical system that transitions between basins of attraction characterized by coherent activity in small groups of brain regions, and that the strength of these attractors depends on the cognitive computations being performed.\end{abstract}

\begin{introduction}
  
The human brain is a complex dynamical system comprised of billions of neurons that continuously communicate with one another.  Although the vast number of processing units challenges exact prediction of single neuron activity, recently developed statistical models reveal a characteristic meso-scale structure whereby sets of larger-scale brain regions display coherent activity at rest. These sets form putative functional modules characterized by locally dense functional connectivity, and include the default mode, salience, attention, fronto-parietal, cingulo-opercular, motor, visual, auditory, and subcortical systems \cite{salvador2005,meunier2009,yeo2011,power2011}. Interestingly, although within-module functional connectivity is in general higher than between-module functional connectivity, these patterns fluctuate dynamically over short periods of time \cite{ma2014dynamic, kiviniemi2011sliding, watanabe2013pairwise}, both at rest and during task performance \cite{cole2014,mattar2015,bassett2011,bassett2013,bassett2015,braun2015}. Yet, fundamental insights into the mechanisms or rules by which modules interact with one another over time have remained elusive \cite{mattar2015}.

One potential route towards a mechanistic theory of brain network dynamics is to consider probabilistic models that were originally developed in the field of statistical mechanics.  Pair-wise maximum entropy models (MEM), for example, have proven very useful in estimating and predicting spiking activity in neurons \cite{shlens2006structure}, local field potentials from neuronal assemblies \cite{tang2008maximum}, and blood oxygen level dependent signals (BOLD) from brain regions using functional magnetic resonance imaging (fMRI) \cite{watanabe2013pairwise,watanabe2014network,watanabe2014energy}.  When a pair-wise MEM accurately fits empirical data, it implies that the observed activation pattern can be described as a combination of each unit's independent activation rate plus the units' joint activation rates. When a pair-wise MEM does not accurately fit empirical data, it implies that higher order interactions (such as triplets) or nonlinearities contribute to the observed dynamics. Importantly, pair-wise MEMs can be used to infer an \textit{energy} landscape of brain activity during task performance, including the presence of common brain states (attractors, or basins on the energy landscape), as well as the paths or trajectories along which the brain moves as it transitions from one basin to another. Importantly, these inferences have proven useful in predicting individual differences in human perception and behavior \cite{watanabe2014energy}.

In traditional applications of MEMs to neurophysiological data, a brain state is defined as a pattern of activity across brain regions (or similarly, a neural state is defined as a pattern of neural activity across neurons). However, these notions of brain state are agnostic to the patterns of communication or synchronization linking brain regions, and therefore are unable to address the question of how one pattern of coherent activity could evolve into another pattern of coherent activity. To address this question, we explicitly define a \emph{network} state as the pattern of module allegiance across brain regions, and we use this definition to examine transitions between network states.  Specifically, we construct a time-dependent network by linking 10 regions of interest by the low frequency (0.06--0.19 Hz) wavelet coherence between their time series in a given time window. We use a community detection algorithm to identify groups of brain regions that show stronger coherence with one another than they do to other groups. We refer to these groups as network communities, and we fit the MEM to each ROI's time series of the state of co-occurrence in the same community with other ROIs. This approach enables us to identify network states that form local energy minima, as well as features of the energy landscape surrounding these minima. More generally, this approach highlights the dynamic functional roles that different ROIs play in network states and the transitions between them.

\begin{figure*}[t]
\centering
\includegraphics[width=.6\linewidth]{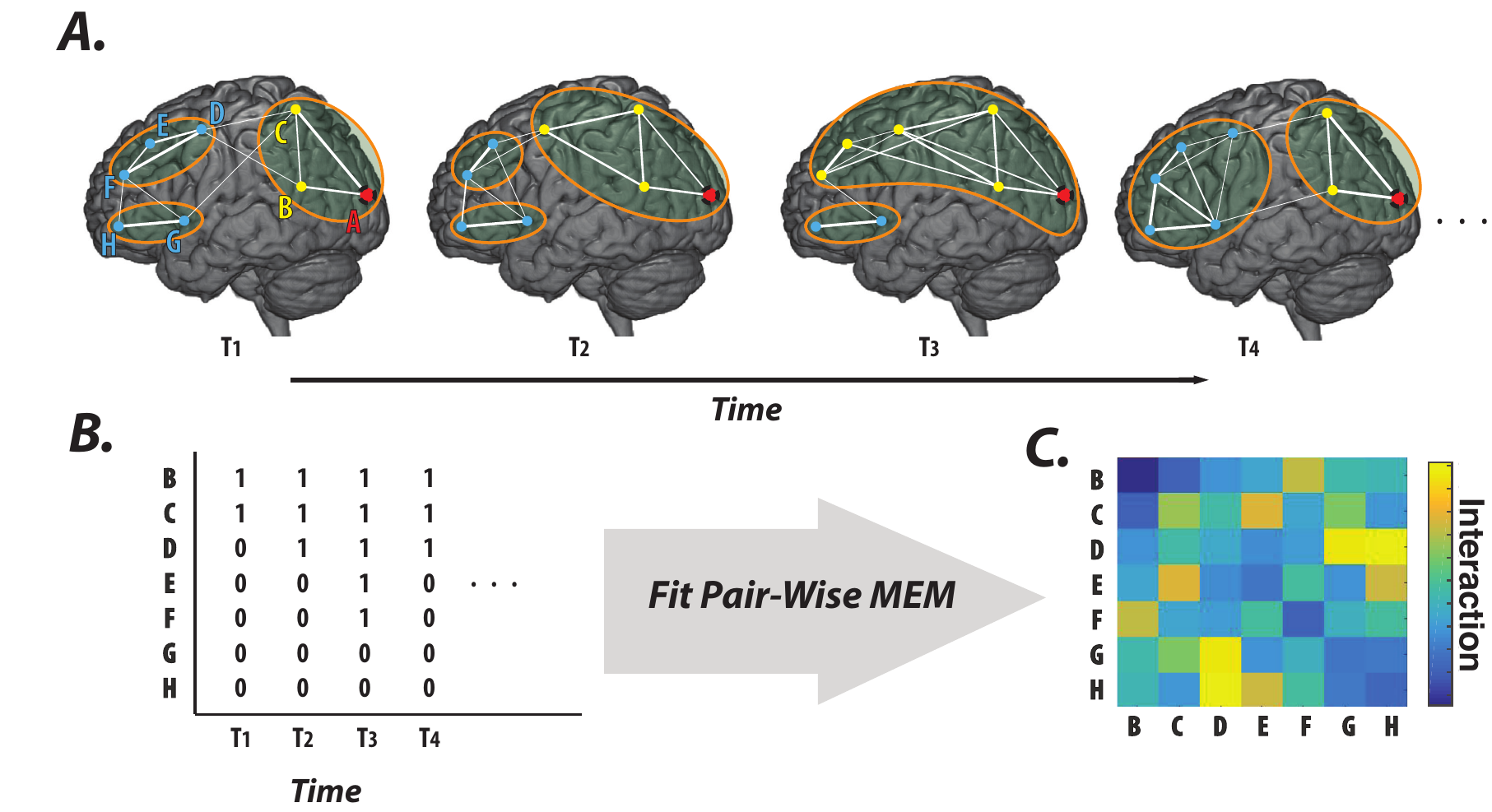} 
\caption{\textbf{Schematic of local community allegiance energy landscape estimation.}. \emph{(A)}  
Fluctuations of the strength of functional connectivity between brain regions over time manifests into reconfiguration of the brain's functional modules. \emph{(B)} The state of the community allegiance of a single region (e.g. node A) with the rest of the brain regions ( '1' when both pairs appear the same community (yellow nodes) and '0' otherwise (blue nodes)) are then used to establish local functional module allegiance states. \emph{(C)} We fit a MEM using these allegiance state vectors and estimate the functional interaction strength between brain regions to construct an energy landscape of regional community allegiance states.}
\label{fig:Intro_Fig}
\end{figure*}
 
Our results reveal the presence of local minima on the energy landscape, many of which are characterized by the activation of a single community. Interestingly, different ROIs show different patterns of membership to these single community states. Visual, attention, sensorimotor, and subcortical ROIs tend to form a single functional community (Class-I). The remaining ROIs form the putative executive control network (Class-II) and the putative default mode and salience network (Class-III). To further study these dynamics, we modeled the transitions of single community states over the landscape of the states's energy via a random walk process. Our numerical simulation of basin transitions using an MCMC random walk predict empirical probabilities of state transitions with high fidelity for Class-I and Class-III ROIs, and with lower fidelity for executive control (Class-II) ROIs. In addition, empirically the executive control ROIs also display higher entropy energy landscapes, linking diverse state classes, and utilizing uniform transition probabilities across basins. These features support the unique role of executive control regions in diversifying the brain's dynamic functional repertoire across many cognitive processes via their rich and flexible dynamic functional fingerprint. 
 
\end{introduction}

\begin{methods}

\subsection*{Participants}
Twenty participants (nine female; ages 19--53 years; mean age = 26.7 years) with normal or corrected vision and no history of neurological disease or psychiatric disorders were recruited for this experiment. All participants volunteered and provided informed consent in writing in accordance with the guidelines of the Institutional Review Board of the University of Pennsylvania (IRB \#801929).
\subsection*{Human fMRI Data collection}

Magnetic resonance images were obtained at the Hospital of the University of Pennsylvania using a 3.0 T Siemens Trio MRI scanner equipped with a 32-channel head coil. T1-weighted structural images of the whole brain were acquired on the first scan session using a three-dimensional magnetization-prepared rapid acquisition gradient echo pulse sequence (repetition time (TR) 1620 ms; echo time (TE) 3.09 ms; inversion time 950 ms; voxel size 1 mm $\times$ 1 mm $\times$ 1 mm; matrix size 190 $\times$ 263 $\times$ 165). A field map was also acquired at each scan session (TR 1200 ms; TE1 4.06 ms; TE2 6.52 ms; flip angle $60^{\circ}$; voxel size 3.4 mm $\times$ 3.4 mm $\times$ 4.0 mm; field of view 220 mm; matrix size 64 $\times$ 64 $\times$ 52) to correct geometric distortion caused by magnetic field inhomogeneity. In all experimental runs with a behavioral task, T2*-weighted images sensitive to blood oxygenation level-dependent contrasts were acquired using a slice accelerated multiband echo planar pulse sequence (TR 2,000 ms; TE 25 ms; flip angle $60^{\circ}$; voxel size 1.5 mm $\times$ 1.5 mm $\times$ 1.5 mm; field of view 192 mm; matrix size 128 $\times$ 128 $\times$ 80). In all resting state runs, T2*-weighted images sensitive to blood oxygenation level-dependent contrasts were acquired using a slice accelerated multiband echo planar pulse sequence (TR 500 ms; TE 30 ms; flip angle $30^{\circ}$; voxel size 3.0 mm $\times$ 3.0 mm $\times$ 3.0 mm; field of view 192 mm; matrix size 64 $\times$ 64 $\times$ 48). 

\subsection*{fMRI Preprocessing}

We preprocessed the resting state fMRI data using FEAT (FMRI Expert Analysis Tool) Version 6.00, part of FSL (FMRIB's Software Library, www.fmrib.ox.ac.uk/fsl). Specifically, we applied: EPI distortion correction using FUGUE ~\cite{jenkinson2004improving}; motion correction using MCFLIRT ~\cite{jenkinson2002improved}; slice-timing correction using Fourier-space timeseries phase-shifting; non-brain removal using BET \cite{smith2002fast}; grand-mean intensity normalization of the entire 4D dataset by a single multiplicative factor; highpass temporal filtering (Gaussian-weighted least-squares straight line fitting, with sigma=50.0s).

Nuisance timeseries were voxelwise regressed from the preprocessed data. Nuisance regressors included (i) three translation (X, Y, Z) and three rotation (Pitch, Yaw, Roll) timeseries derived by retrospective head motion correction ($R=[X,Y,Z,pitch,yaw,roll]$), together with expansion terms ($[R R^{2} R_{t-1} R_{t-1}^{2}]$), for a total of 24 motion regressors~\cite{friston1996movement}); (ii) the five first principal components calculated from timeseries derived from regions of non-interest (white matter and cerebrospinal fluid), using the anatomical CompCor method (aCompCor) \cite{behzadi2007component} and (iii) the average signal derived from white matter voxels located within a 15mm radius from each voxel, following the ANATICOR method \cite{jo2010mapping}. Global signal was not regressed out of voxel time series due to its controversial application to resting state fMRI data \cite{murphy2009impact, saad2012trouble, chai2012anticorrelations}. Finally, the mean functional image and the 125-scale Lausanne parcellation template \cite{cammoun2012mapping} were coregistered using Statistical Parametric Mapping software (SPM12; Wellcome Department of Imaging Neuroscience, www.fil.ion.ucl.ac.uk/spm) in order to extract ROIs' mean timeseries. 

\subsection*{ICA-Informed Identification of Regions of Interest }

We used group-ICA (GIFT toolbox \cite{calhoun2001method}) to identify ten large-scale intrinsic connectivity networks (ICN) characteristic of the resting state. Computational considerations preclude us from studying a larger number of networks in the context of the maximum entropy model approach. Next, we identified the regions of interest (ROIs) by choosing the ROI from a commonly used anatomical atlas (the 125 scale Lausanne parcellation \cite{cammoun2012}) in which we observed the peak activation of an ICA component. A list of identified ICNs and their corresponding Lausanne atlas ROI are provided in Table ~\ref{tab:table1}.  In supplementary figure SI7, we also discuss the effect of anatomical versus functional parcellation methods on our results.

\begin{figure*}
\centering
\includegraphics{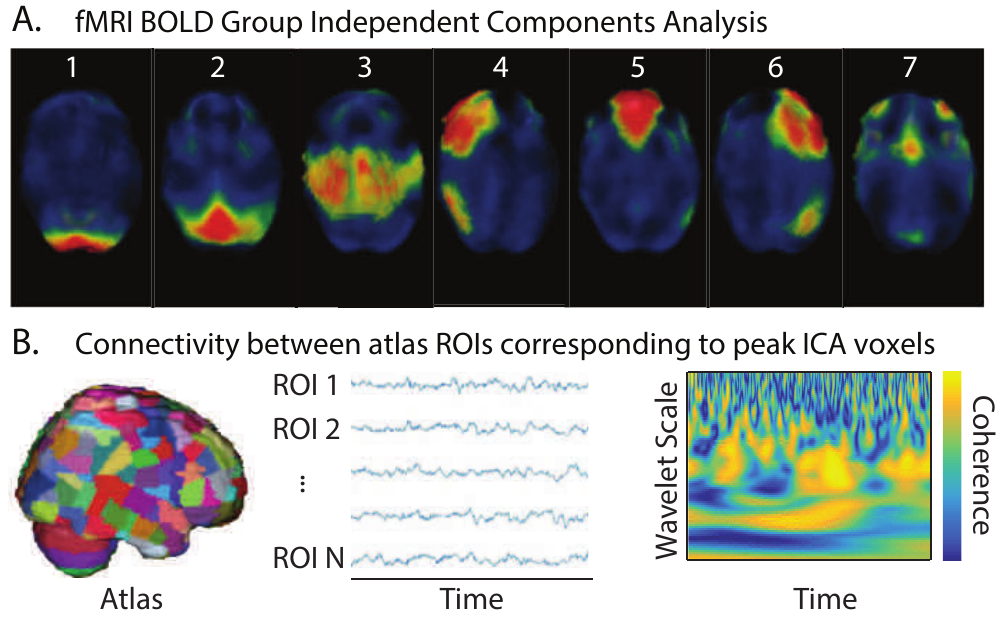}
\caption{\textbf{Schematic of Methods}. \emph{(A)} We used a group-ICA decomposition to distill fMRI resting state BOLD into $N (=10)$ components representing putative baseline functional networks. For each ICN, we identified the voxel with the peak expression of that component. \emph{(B)} Using the Lausanne 125 scale template (234 ROIs) \cite{cammoun2012}, we determined the atlas region corresponding to the peak expression of each component. After extracting BOLD time series from each ROI, we estimated the functional connectivity between pairs of ROIs using the wavelet coherence in the frequency interval $0.19--0.06Hz$ \cite{bassett2011}.}
\label{fig:Figure1}
\end{figure*}

For computational reasons which we will discuss in more details shortly, we only included ROIs from the left hemisphere for the majority of components. The one exception to this rule was the right executive control network (RECN). Prior work has demonstrated that the putative fronto-parietal executive control network is identified as two separate bilateral ICNs ~\cite{shirer2012decoding,laird2011behavioral}, which separately play critical roles in executive function through their dynamical interactions with the DMN and attention systems \cite{andrews2014default,menon2010saliency}. Thus we included time series from both the right and left executive control networks in all following analysis.

\begin{table}
\centering
\caption{Regions of interest and their corresponding ICN }
\begin{tabular}{lr}
 ICN & Lausanne ROI (scale 125)  \\
\midrule
1. Visual & 189 Cuneus.1  \\
2. Dorsal Attention (Attn) & 184 Precuneus.1  \\
3. Sensory/Motor (SM) & 147 Precentral.3  \\
4. Basal Gangla/Thalamus (BG)  & 228 Caudate  \\
5. Left Executive Control Network (LECN)  & 128 Rostral middle frontal.2   \\
6. Right Executive Control Network (RECN)  & 15 Rostral middle frontal.2  \\S
7. Rostral Middle Frontal Cortex (rmFC)  & 130 Rostral middle frontal.4  \\
8. Dorsomedial Prefrontal Cortex (dMPFC)  & 135 Superior frontal.3  \\
9. Default Mode Network (DMN) & 122 Medial orbitofrontal.1   \\
10. Salience  & 124 Pars triangularis.1   \\

\bottomrule
\end{tabular}
\label{tab:table1}
\end{table}

\subsection*{Functional Network Construction}

Following prior work \cite{bassett2011}, we estimated the dynamic functional connectivity between all pairs of ROIs using wavelet coherence (WTC \cite{grinsted2004}). In supplementary figure SI4, we show that we observe two distinct bands of high WTC: $0.64--0.2Hz$ and $0.19 -- 0.06 Hz$. In the main manuscript, we focus on the $0.19 -- 0.06 Hz$ band due to known sensitivity to underlying neural activity, and we relegate discussion of the higher frequency band to the supplement. WTC amplitudes were averaged over all frequencies within the selected band to construct the timecourse of the band-passed WTC for each pair of ROIs resulting in a total of T (=1190 (TRs) $\times$ 20 (subjects)  $\times$ 4 (runs) = 95200) unique functional connectivity patterns, which we represent in $N \times N$ adjacency matrices $\mathbf{A}$(see Fig.~\ref{fig:Figure1}).

\subsection*{Community Detection and Module Allegiance Estimation}

In the maximum entropy framework, it is critical that data points are as temporally distinct from one another as possible. In the context of our study, this requires that we reduce the dependence of community structure in the neighboring time slices. To do so, we identified the community structure \cite{fortunato2010,porter2009} of the each time slice adjacency matrix independently using a Louvain-like \cite{blondel2008} locally greedy heuristic algorithm to maximize the modularity quality function \cite{newman2006} with a structural resolution parameter of $\gamma =1$ \cite{bassett2013robust}.The method partitions ROIs into communities based on the optimization of the following function:
\[
Q_{0} = \sum_{ij} \left [ A_{ij}-\gamma P_{ij} \right ]\delta \left ( g_{i}, g_{j} \right )
\]

where $\mathbf{A}$ is a weighted adjacency matrix, ROI$\textit{i}$ and  $\textit{j}$ are assigned respectively to community $g_{i}$ and $g_{j}$, the Kronecker delta $\delta \left ( g_{i}, g_{j} \right ) = 1$  if $g_{i}= g_{j}$  (and zero otherwise), $\gamma$  is the structural resolution parameter, and $P_{ij}$  is the expected weight of the edge between ROI$\textit{i}$ and $\textit{j}$ under some null model. We used the Newman-Girvan null model \cite{girvan2002}

\[
P_{ij} = \frac{k_{i}k_{j}}{2m}
\]

where  $\textstyle k_{i} =\sum_{j} A_{ij}$ is the strength of ROI $\textit{i}$ and $\textstyle m = \frac{1}{2} \sum_{ij} A_{ij}$. In a nutshell, this method partitions the ROIs into groups such that the total connection strength within communities is more than expected in the null model. 

Importantly, the algorithm we use is a heuristic that implements a non-deterministic optimization \cite{good2010}. Consequently we repeated the optimization 100 times \cite{bassett2013robust}, and we report results summarized over those iterations by forming a module allegiance matrix \cite{bassett2015,mattar2015}.

The allegiance matrix for each time slice represented the probability that ROI $\textit{i}$ and $\textit{j}$ were assigned to the same community over all iterations of the community detection algorithm (Fig.~\ref{fig:Figure2a}). For use in the maximum entropy model, we binarized the allegiance matrix by subtracting a random null model allegiance matrix from the original allegiance matrix: elements in the binarized allegiance matrix were 1 when the true allegiance was greater than the null, and 0 otherwise \cite{bassett2013robust}. The null allegiance matrices were generated by shuffling the ROI community assignments for each individual time point uniformly at random.

\begin{figure*}[tbhp]
\centering
\includegraphics[width=.7\linewidth]{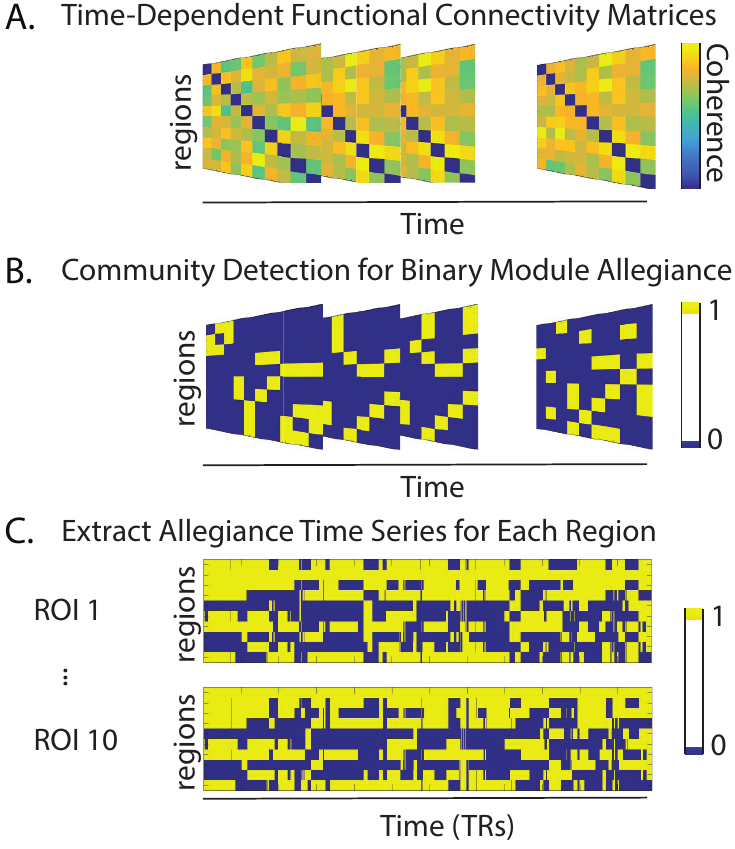}
\caption{\textbf{Schematic of Methods for Extracting Dynamic Module Time Series.} \emph{(A)} We represent the T (=1190 (TRs) $\times$ 20 (subjects)  $\times$ 4 (runs) = 95200) unique functional connectivity patterns as $N \times N$ adjacency matrices $\mathbf{A}$. \emph{(B)} Using community detection, we extract putative functional modules at each TR, and use a statistical comparison to a random null model to determine a region's binary module allegiance. More specifically, we identify the  community organization of ROIs and calculate the probability of ROI pairs' congruent community allegiance for each time-point. The ROI paris with higher than expected (via permutation tests) congruent community allegiance were thresholded to generate binarized pairwise allegiance matrices. \emph{(C)} We reformat these data to separately store the allegiance time series of each ROI, which codes is co-allegiance with other ROIs to the same community (values of 1) as a function of time (TR).
\label{fig:Figure2a}}
\end{figure*}

\subsection*{Maximum Entropy Model Fitting}

Here we hypothesize that the brain transitions between different functional community states. To obtain an unbiased estimate of these states and their probabilities, we fit a pair-wise maximum entropy model. The principle of maximum entropy states that when estimating the probability distribution, given the constraints, one should find the distribution that maximizes the uncertainty (i.e., entropy). Choosing any other distribution that lowers the entropy would assume information that we do not possess; therefore the only reasonable distribution is the maximum entropy distribution. Fitting the MEM entails tuning the first and second-order interaction parameter between regions so that the predicted activation rates and co-activation rates match that of the empirically observed values. An accurate pair-wise MEM fit suggests that the observed dynamics of the communities can be simply explained as a combination of each region's independent activation rate plus the region's joint activation rates. In other words, the MEM allows us to establish a mechanistic model of brain functional dynamics as a probabilistic process shaped by intrinsic relationships between brain regions.

We fit the pairwise MEM to the binarized community allegiance of ROI pairs (similar to \cite{watanabe2013pairwise}). To reduce the size of the state space and therefore ensure less error in our estimates, we fit the MEM to each row of the allegiance matrix independently for each ROI, effectively reducing the dimensionality to N-1 (= 9) and reducing the total number of possible congruent community membership states to $2^{9} = 512$. For ROI $\textit{i}$ at time $\textit{t}$ the congruent community membership state is defined as  $\textstyle V^{t}=\left [ \sigma_1^t , \sigma_2^t ,..., \sigma_{i-1}^t, \sigma_{i+1}^t,...,, \sigma_N^t \right ]$, where $\sigma_{j}^t$ is the binarized community allegiance of $\textit{i}$ and $\textit{j}$ at time point $\textit{t}$ ('1' for congruent community membership and '0' otherwise), and N is the total number of ROIs (= 10). For ROI $\textit{i}$ the empirical congruent community membership rate of ROI $\textit{j}$, $\left \langle \sigma_j  \right \rangle$, is given by $\textstyle (\frac{1}{T}) \sum_{t=1}^T \sigma_j^t$, where T is the number of time slices, which in this case is equal to the number of TRs. Likewise the empirical pairwise congruent community membership rate of ROIs $\textit{j}$ and $\textit{l}$, $\left \langle \sigma_j , \sigma_l \right \rangle$, is defined as $\textstyle (\frac{1}{T}) \sum_{t=1}^T \sigma_j^t  \sigma_l^t$ .

 Here our only constraints were that the model $\left \langle \sigma_j  \right \rangle_{m}$ and $\left \langle \sigma_j\sigma_l  \right \rangle_{m}$ matched the empirical values of  $\left \langle \sigma_j  \right \rangle$ and $\left \langle \sigma_j \sigma_l  \right \rangle$ respectively. It is known that given these constraints the probability distribution that maximizes the entropy is the Boltzman distribution \cite{jaynes1957information} 
\begin{equation}
P(V_{k})=e^{-E(V_{k})}/ \sum_{q=1}^{2^{N-1}} e^{-E(V_{q})}
\label{dyn}
\end{equation}
where $P(V_{k})$ is the probability distribution of $\textit{kth}$ state $V_{k}$, and $E(V_{k})$ the energy of that state given by
\begin{equation}
E(V_{k}) = -  \sum_{j=1}^{N-1} h_{j} \sigma_{j}(V_{k}) -\frac{1}{2} \sum_{j=1}^{N-1} \sum_{l=1, j\neq l}^{N-1} j_{jl}\sigma_{j}(V_{k})\sigma_{l}(V_{k})
\end{equation}
 where $\sigma_{j}(V_{k})$ is the value of $\sigma_{j}$  for state $V_{k}$, $h_{j}$ represents the expected base allegiance of ROI $\textit{j}$ (with respect to ROI $\textit{i}$) in isolation, and $j_{jl}$ represents the functional interaction between ROI $\textit{j}$ and $\textit{l}$. Fitting the MEM entails iterative adjustment of $h_{j}$ and $j_{jl}$ with a gradient ascent algorithm (similar to \cite{watanabe2014energy}) 
until the empirical $\left \langle \sigma_j \right \rangle$ and $\left \langle \sigma_j  \sigma_l  \right \rangle$ values approximately match the model $\textstyle\left \langle \sigma_j   \right \rangle_{m} = \sum_{q=1}^{2^{n-1}} \sigma_{j}(V_{q})P(V_{q})$ and $\textstyle\left \langle \sigma_j  \sigma_l  \right \rangle_{m} = \sum_{q=1}^{2^{n-1}} \sigma_{j}(V_{q})\sigma_{l}(V_{q}) P(V_{q})$.  In depth analysis of the goodness of MEM fits (provided in SI1), allows us to conclude that the pair-wise MEM can account for a large portion of the observed functional module dynamics. Nevertheless, higher-order and/or nonlinear interactions likely contribute to smaller yet non-negligible portions of the observed brain dynamics.

\begin{figure*} 
\centering
\includegraphics[width=0.65\linewidth]{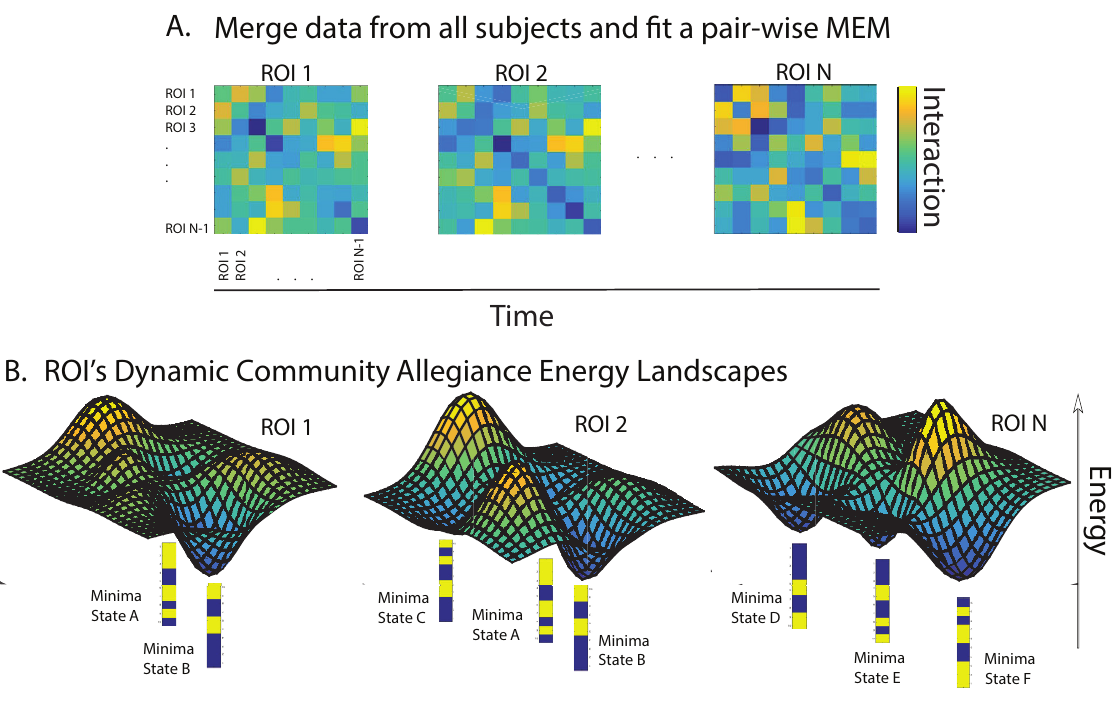}
\caption{\textbf{Schematic of Maximum Entropy Model of Brain Network Dynamics.}  \emph{(A)} Accurate fitting of a MEM requires large amounts of data. We therefore combined data across all subjects before fitting a pairwise MEM to each ROI's pattern of allegiance to functional modules. This fitting procedure produced an estimated interaction matrix for each ROI and each TR; colors indicate the strength of each element of the interaction matrix $J_{ij}$. \emph{(B)} From the interaction matrices, we defined and characterized energy landscapes of the local community dynamics. Color indicates energy, with yellow indicating high energy and dark blue indicating low energy. Each minimum within each landscape is accompanied by an example network state, as defined by a binarized pattern of module allegiance (yellow indicating congruent module allegiance and dark blue indicating incongruent module allegiance) with other ROIs.
\label{fig:Figure2b}}
\end{figure*}

\subsection*{Defining an Energy Landscape}

The energy landscape for each ROI is defined separately by the network of congruent community membership states $V_{k}$ and their corresponding energy values $E(V_{k})$. In this landscape the adjacent states' vectors are only one hamming distance apart, which means that all ROIs except one have the same binary values as in adjacent states' vectors. \cite{fninf.2014.00012} 

An interesting question to ask of this landscape is the location and nature of energy minima. To address this question, we exhaustively searched the entire landscape using a steep search algorithm to find the states with energies lower than all their neighboring states (i.e. local minima). Next in order to identify the states that belong to the basin of each local minima we first start at a given state $V_{k}$ (one of the $2^{N-1} (=2^{9}=512)$ possible states) and iteratively move to the neighboring state $V_{w}$ in the landscape if $E(V_{w}) < E(V_{k})$. We continue tracing out this path until we reach a local minima state where no neighboring states exist with smaller energy values (similar to \cite{watanabe2014energy,watanabe2013pairwise}). We consider this final state $V_{k}$ the basin state of the local minima. We also define the basin size of that local minima state as the fraction of the number of basin states to the total number of possible states.

We were next interested in understanding the predicted barriers between states. We estimated the energy barrier opposing the transition between all the local minima states in the following way: (1) We removed the state (node) with the highest energy from the energy landscape along with the edges connecting that state to its neighbors. (2) We assessed whether each pair of local minima were connected by a path in the reduced landscape. We repeated steps (1) and (2) until we found the saddle state where removing the highest energy node disconnects one or more local minima from the rest of the landscape. We continued this process until we obtained a reduced landscape where all of the local minima are isolated and we identified all saddle states. (3) We calculated the symmetric energy barrier \cite{zhou2011random} between all pairs of minima states as the minimum of $\textstyle \left [ E^{S}(V_{k},V_{w})-E(V_{k}),E^{S}(V_{k},V_{w})-E(V_{w}) \right ]$, where $E^{S}(V_{k},V_{w})$ is the energy of the saddle point between minima states $V_{k}$, $V_{w}$ and $E(V_{k})$ and $E(V_{w})$ are the energies at these states, respectively. If the energy barrier between to minima states was high then the model predicts that the rate of transition between them is low, at least in one direction \cite{watanabe2013pairwise}. 
We also calculated the asymmetric energy barrier between minima states $V_{k}$, $V_{w}$ as $\textstyle E^{S}(V_{k},V_{w})-E(V_{k})$ and $E^{S}(V_{k}$, $V_{w})-E(V_{k})$, where the former indicated the $V_{k}\rightarrow V_{w}$ and  the latter the $V_{w}\rightarrow V_{k}$ energy barriers. Overall, our results did not show any relationship between estimated energy barriers (symmetric and asymmetric) between local minima states and the empirical basin transition probabilities. We speculate that the energy barrier mimics the basin transition probabilities only when the basins are smooth and funnel-like. Thus unaccounted factors such as the shape (e.g., roughness) of the basins may contribute more to the observed basin transition probabilities than the energy barriers between the basins.

We speculate that other unaccounted factors such as the shape of the basins may contribute more to the observed basin transition probabilities than the energy barriers between the basins.

\subsection*{Simulation of State Transitions}

To better understand the dynamic patterns of functional communities at rest, we simulated these dynamics as a random walk process over the estimated local energy landscapes using a Markov chain Monte Carlo with Metropolis-Hastings (MCMC) algorithm \cite{metropolis1953equation,hastings1970monte,zhou2011random}.  In this model, local community allegiance state $V_{i}$ is allowed an isometric transition to one of $N-1$ neighboring state with uniform probability. Next the actual transition from $V_{i}$ to $V_{j}$ occurs with probability $\textstyle P_{ij}=min\left [ 1,e^{E(V_{i})-E(V_{j})} \right ]$. For each ROI, we repeated a $4\times10^{7}$ step (plus  $3\time10^{4}$ initial steps) walk with randomly chosen initial states $2\time10^{4}$ times. Next, we removed the initial steps to ensure independence of results from the initial conditions and decreased the sampling rate by 500 to reduce the correlation between the samples. Since each state in the energy landscape belongs to the basin of a single local minima, we can construct a trajectory of local minima states' basin transitions from the down-sampled state transitions patterns. Comparing the empirical and simulated basin transition probability patterns allows us to evaluate the resemblance of the proposed random walk model's dynamics to that of the brain.

\subsection*{A note on Statistics and Computational Considerations}

The fit of the maximum entropy model to small datasets can be subject to a sampling bias. More exactly, the estimation of the entropy suffers severely from downward bias \cite{treves1995upward}  such that the estimated entropy from the observations is lower than the actual entropy of the underlying model ~\cite{macke2011biased}. Ultimately the amount of data needed to accurately fit the model is exponential to the number of ROIs. Therefore even a small number of ROIs (on the order of ten) requires access to extremely large datasets. This issue of computational complexity is especially critical for fMRI data where the slow sampling rate prohibits collection of individual subject dataset with a large number of observations. Consequently, in this manuscript we focus on group concatenated multiband data (with a quarter-second TR) and do not discuss results from individual subjects. In supplementary figure SI1, we briefly discuss the model fit at the subject- and group-level where we demonstrate that the goodness of fit and accuracy of the model drops considerably when considering individual subjects as opposed to the group.

The total number of partitions of an $n$-element set is the Bell number $n$ $B_{n}$~\cite{grimaldi2006discrete}. In combinatorial mathematics, the Bell numbers count the number of partitions of a set. Thus the total number of possible community states of the 10 ROIs equals 115975. To accurately model this large number of states would require a large amount of data. However, local analysis (through the lens of a single ROI) of congruent community allegiance deals with a much smaller state space of $2^{N-1} ( = 512)$. At this level, it is computationally feasible to fit a MEM with a relatively large multiband fMRI dataset such as the one we use here. 

\end{methods}

\begin{results}

 \paragraph{Distillation of Drivers of Resting State Dynamics}

Maximum entropy models are optimally constructed to fit patterns of interactions between \emph{relatively few} brain regions. We therefore sought to distill the drivers of resting state dynamics to a few well-chosen regions of interest. Specifically, in resting state fMRI data acquired from 20 healthy adult individuals in a multiband imaging sequence, we extract 10 regions of interest in a data-driven fashion as centroids of independent components (see Methods). These regions include the cuneus, precuneus, precentral gyrus, caudate, right and left rostral middle frontal cortex, dorsomedial prefrontal cortex, medial orbitofrontal cortex, and pars triangularis (see Table~\ref{tab:table1}). We use these regions as proxies of their respective cognitive systems, spanning visual, dorsal attention, sensorimotor, basal ganglia, executive control, dorsomedial prefrontal cortex, default mode, and salience systems, respectively. For explicit maps of each independent component, and the representative region chosen, see SI3. 

\paragraph{Maximum Entropy Model of Network States}

Our goal is to understand how the brain transitions between network states. We focus our attention on the transitions characterized by changes in the community structure of the network, or the organization of putative functional modules. This focus is motivated by a growing literature demonstrating (i) the presence of network communities at rest, which map on to known cognitive systems \cite{salvador2005,meunier2009,power2011,yeo2011}, and (ii) changes in the integration or segregation of these communities during task performance \cite{bassett2011,bassett2013,cole2014,bassett2015,braun2015}. Based on these emerging lines of research, we define network states based on regions' congruent community allegiance. Specifically, we fit a MEM to the binarized community allegiance probability of region $\textit{i}$ where $(i= 1,2,..,N)$ with the other $9$ ROIs. This approach assesses the community-based interactions between a single ROI and all others, thus significantly reducing the space of possible states (from 115975 possible global community states to only  $2^{N-1} ( = 512)$ local single community states) and thereby increasing model accuracy \cite{grimaldi2006discrete}.

\paragraph{Local Minima in the Brain's Functional Energy Landscape.}

After fitting the MEM to regional congruence in community allegiance across all subjects, we characterized the resultant energy landscapes of all ROIs. We identified 3--5 local minima states from the landscape of a single ROI, for a total of $M=25$ unique local minima states across all ROIs. We note that each state represents the set of brain regions that are commonly allied together in a single community (see Fig.~\ref{fig:Figure3}A). Because ROIs with strong functional interactions are expected to display congruent membership in local minima states, we performed hierarchical clustering on the pattern of ROI allegiance to local minima states. Interestingly, we observed that ROIs divided neatly into three separate classes. Class-I was composed of occipital, parietal, and subcortical ROIs in the visual, attention, sensorimotor, and basal ganglia systems. Class-II was composed of fronto-parietal, and frontal ROIs in the right and left executive control network and the rostral middle frontal systems. Finally, Class-III was composed of medial and opercular ROIs in the dorsal medial prefrontal, default mode, and salience systems.

\begin{figure*} 
\centering
\includegraphics[width=0.8\linewidth]{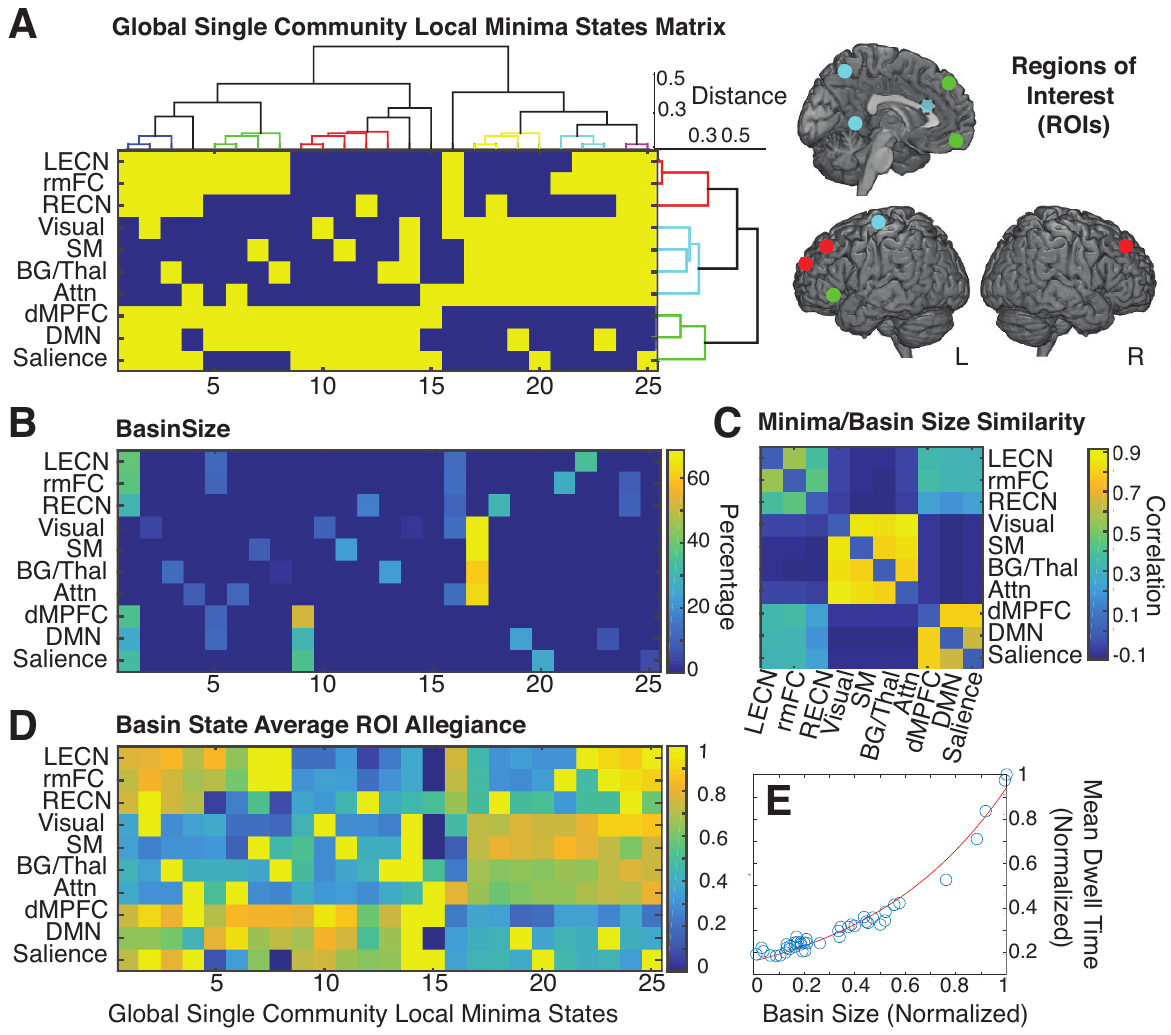}
\caption{\textbf{Local Minima in the Brain's Functional Energy Landscape.} \emph{(A)} We identified $M=25$ unique local minima states characteristic of the time series of a region's allegiance to putative functional modules over time. Here, each state represents the set of regions that are commonly allied together in a single community. Using hierarchical clustering, we identified three classes of ROIs with similar patterns of community allegiance across these local minima. Class-I (green), Class-II (blue), and Class-III (red) branches are shown in the dendrogram on the right side of the matrix and visualized on the brain images as ROIs. Using the same clustering technique applied to the matrix transpose, we identified several classes of minima states with common ROI members, denoted by the colored branches on the top of the matrix. ICN labels and their corresponding Lausanne atlas ROI are provided in Table \ref{tab:table1}. \emph{(B)} In addition to identifying the unique minima states, for each ROI we also calculated the \emph{basin size} of local minima states as the fraction of the number of basin states to the number of total possible states. \emph{(C)} Calculating the Pearson correlation coefficient between any two ROIs' vectors of basin size across minima states revealed groups of ROIs with similar energy landscapes. \emph{(D)} The basin states' average vectors highlight the unifying features of the basin states, i.e. omni-present core ROIs (average value 1), commonly present core ROIs (average value between $0.5$ and $1$) that tend to be among the minima states' member ROIs, and finally the ROIs with incongruent membership (average value $<0.5$). \emph{(E)} The average dwell time of minima state's basins (all ROIs combined) grows exponentially with respect to the basins' size. The close exponential fit (red curve) highlights this relationship.}
\label{fig:Figure3}
\end{figure*}

Although we identified a total of 25 unique local minima states across all ROI landscapes, the majority of these states approximated single communities consisting largely of ROIs from one or two identified classes. For example as see in Fig.~\ref{fig:Figure3}A, local minima states 1--3 are characterized by congruent community allegiance of Class-II and Class-III ROIs. Hierarchical clustering of local minima states based on their similarity (as measured by the Hamming distance between the local minima states) underlines several groups of very similar local minima states, where each group is characterized by a few common ROIs. Interestingly, these results highlight the tendency of Class-I and Class-III ROIs to display incongruent (or dissimilar) community allegiance, whereas Class-II ROIs form communities with ROIs across both classes. This observation provides converging evidence of the distinctive role of the putative executive control regions in diversifying the brain's dynamic functional repertoire across many cognitive process via their functional fingerprint.

\paragraph{Basins surrounding local energy minima}

In the previous section, we identified minima of the energy landscape underpinning module dynamics, and we further described the clustering of these minima states into groups with similar regional profiles of allegiance to dynamic modules.  These minima were located at the locally least-energetic position on the energy landscape. However, it is also intuitively of interest to study the group of states neighboring each local minimum since the shape of the surrounding low-energy basin effects the state transition dynamics over the basin.  The brain states are expected to rapidly converge to the local minima in a smooth, steep, funnel-like basin, whereas over the larger rough basins the systems displays frustrated dynamics; i.e. it is difficult to reach local minima because of the large number of peaks and troughs. To study this larger group of states, we defined \emph{basin states} of local minimum \textit{A} to be any state from which one can travel down the steepest gradient of the energy landscape (using a steep search algorithm) to reach the local minimum \textit{A} (see Methods for details). Using this definition, we estimated any basin's size as the ratio of the number of basin states to the total number of possible states.  We observed that the largest basin -- making up to $>60 \%$ of the total state space of Class-I ROIs -- surrounded minima state \#17, which was characterized by congruent allegiance of visual, sensorimotor, basal ganglia, and attention systems (see Fig.~\ref{fig:Figure3}B). The large size of this basin suggested that these systems had stronger than average intrinsic functional interactions, perhaps due to the ongoing visual fixation that is characteristic of the resting state with eyes open.

To better understand the anatomical drivers of observed module dynamics, we next studied which brain regions contributed most to basin states surrounding each minimum.To address this goal, we began by counting the number of times that each basin state appeared in a single region's profile. Using this information, we defined for each ROI the number of basin states seen by that ROI as a fraction of the total number of possible states. In other words, associated with each ROI was a vector that represented the fraction of its states that were identified in the basin of each of the minima states. Interestingly, the identified classes of ROIs seemed to show similar profiles, as measured by a Pearson correlation coefficient between any two ROIs' vectors (see Fig.~\ref{fig:Figure3}C). Moreover, with the exception of a few states (e.g. states 7 and 8), the ROIs from the same class appear with the same frequency between different basin states; in Fig.~\ref{fig:Figure3}D, this effect is evident through the comparable average allegiance values of ROIs from the same class. 

To understand how local minima states with large basin sizes effect the state transition dynamics, we calculated the \emph{dwell time} of each ROI in a given basin. Intuitively, a dwell time is the length of time in which an ROI remains in a given local minima's basin. We observed that the size of the basin was positively related to dwell time within the basin (see Fig.~\ref{fig:Figure3}E). We note that on average the dwell time of each basin is only a few seconds ($< 10 ~secs$), except for state \#17 which displays twice the average dwell time ($21.46 ~secs$) although the variance is large ($std = 31.41~ secs $). In addition, supplementary Figures SI1-2 also reveal the relatively rough surface of the state \#17 basin, marked by the presence of several prominent peaks and troughs. Together these results suggest that close functional relationships between ROIs (especially between Class-I ROIs) promotes frustrated dynamics over large basins, marked by coherent activity in Class-I ROIs.

\paragraph{The probability of transitioning between basins}

After identifying local minima states and the times spent dwelling within their basin, we next turned to examining the probabilities with which the brain transitioned between single community states to estimate an ROI's transition profile. Because each time point of the subjects' dataset is often associated with two or more single community states, we estimated the probability of transitioning from one basin to another, separately for each ROI's energy landscape (see Fig.~\ref{fig:Figure6}). For Class-I ROIs that include occipital, parietal, and subcortical areas in the visual, attention, sensorimotor, and basal ganglia systems, we observed that the basin of minima state \#17 (characterized by the largest basin size and by congruent allegiance of visual, sensorimotor, basal ganglia, and attention systems) was also the most frequently visited basin. These results once again highlight the strength of this attractor state, which in turn echoes the close functional interactions between Class-I ROIs.  For Class-II ROIs including fronto-parietal and frontal areas in the right and left executive control network and the rostral middle frontal systems, we observed a more uniform distribution of transition probabilities between basins. This more uniform transition probability architecture is also characteristic of Class-III ROIs, which are composed of medial and opercular areas in the dorsal medial prefrontal, default mode, and salience systems. 

To have a better understanding of the fluidity of the state transitions for each ROI, we measured the level of unpredictability (i.e., entropy) of each ROI's community allegiance states. We calculated the entropy of allegiance state probabilities for each ROI separately. The significantly higher entropy of the Class-II ROI landscapes ($p < 0.01$ bootstrap) provides converging evidence of the highly dynamic community organization of ECN ROIs where their single community states more fluidly transition between local minima state basins as highlighted in Fig.~\ref{fig:Figure6} (right). This property of ECN ROIs likely facilitates the adaptation of the brain to the real-time demands of a wide range of cognitive functions by fluidly transitioning between a diverse set of functional community states.

\begin{figure*}
\centering
\includegraphics[width=5.5in]{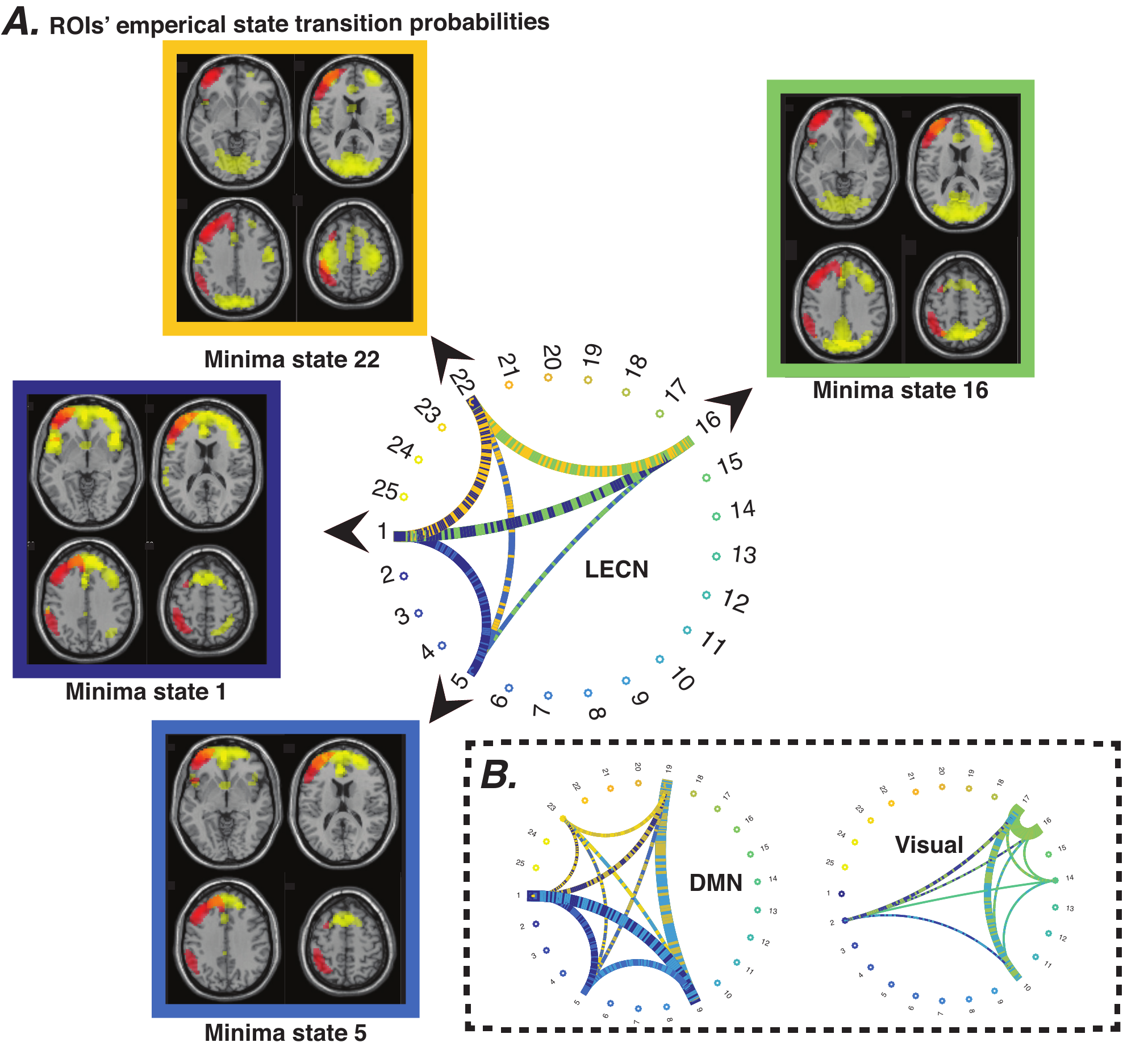}
\caption{\textbf{Empirically Estimated Transition Probabilities Between Basins.}  \emph{(A)} The circular graph represents the empirical basin transition probability pattern of a sample Class-II ROI representing the left hemisphere executive control network (LECN). Each color coded dot along the circumference of the circular graph represents one of the $25$ local minima states. Lines linking local minima states indicate empirically estimated transition probabilities between local minima states's basins; the width of the line is proportional to the empirically estimated probability of that transition. Transition probabilities are separately normalized for ROIs by dividing by the largest transition probabilities calculated for each ROI.  All ICNs with community allegiance congruent with LECN (red overlay) are represented with yellow brain overlays ($z > 1.5$) for all 4 local minima states. \emph{(B)} The empirical basin transition probability pattern for two sample ROIs from Class-I (Visual) and Class-III (DMN). Note that unlike LECN, the basin transition probability pattern of the visual ROI is heavily skewed towards a single state (that is, state 17, which we describe in greater detail in the body of the text). }
\label{fig:Figure6}
\end{figure*}

Next, we asked whether we could use the maximum entropy model results to predict the empirically observed transition probabilities, as a confirmation of the modeling framework. To derive theoretically expected transition probabilities, we simulated the transition dynamics between basins using a Markov Chain Monte Carlo (MCMC) method over the energy landscape of each ROI using the Metropolis-Hastings algorithm (see Methods for details). A secondary benefit of this approach was that it allowed us to identify groups of brain regions whose module dynamics were not well-predicted by a random walk model, and therefore might require consideration of more highly constrained walk dynamics.  In general, we observed that simulated and empirical transition probabilities were positively correlated with one another (average Pearson correlation coefficient $r = 1.09$, $p<0.0002$ except LECN for which $p = 0.02$), offering initial validation of the MEM approach. Interestingly, however, this relationship displayed differential strength across the three classes of ROIs (see as Figure~\ref{fig:Figure5}A and SI2 for more details), indicating that some transition probabilities were less well characterized by pairwise interactions (the assumption of the MEM) and/or by random walks across the underlying energy landscape (the assumption of the MCMC). In particular, Class-I ROIs showed the strongest relationships between the simulated and empirical transition probabilities; Class-III ROIs showed significant but weaker correlations and Class-II in general showed insignificant correlations after FDR correction for multiple comparisons. In the earlier MEM analysis, the functional activity Class-II ROIs were well fit, yet the transition probabilities here find low similarity between the model and empirical data. This suggests that a simple random walk model fails to capture the state transition behavior of putative areas in the executive control networks, which form the bulk of Class-II ROIs.

Finally, we asked whether the empirical transition probabilities could be predicted by simpler statistics drawn from the maximum entropy model (and associated energy landscape) and not requiring the full MCMC modeling approach. In general, we observe a positive relationship between the size of a basin and the transition probability of that basin: the brain tends to transition into and out of large basins. This effect is strongest in Class-I and Class-II ROIs, and weaker for Class-III ROIs (Figure~\ref{fig:Figure5}B). Intuitively, while the basin size is likely to be a strong predictor of transition probability, another important consideration lies in the energy barriers between basins. That is, are two basins separated by a low hill or by a high mountain on the energy landscape? To clarify the relative predictive power of basin size versus barriers between basins, we estimated the energy barrier between pairs of basins by identifying saddle nodes on the energy landscape (see Methods). Importantly, we could estimate these barriers either by considering symmetric transition probability estimates (averaging both transitions into and out of a state), or by considering the full transition probability matrix with small asymmetries. We did not observe any consistent trend linking the size of the energy barrier and the empirically observed transition probabilities (Figure~\ref{fig:Figure5}C), particularly in the case of the asymmetric estimates (Figure~\ref{fig:Figure5}D). These results suggest that module dynamics are best explained by basin size rather than by barriers between basins.

\begin{figure*}
\centering
\includegraphics[width=6in]{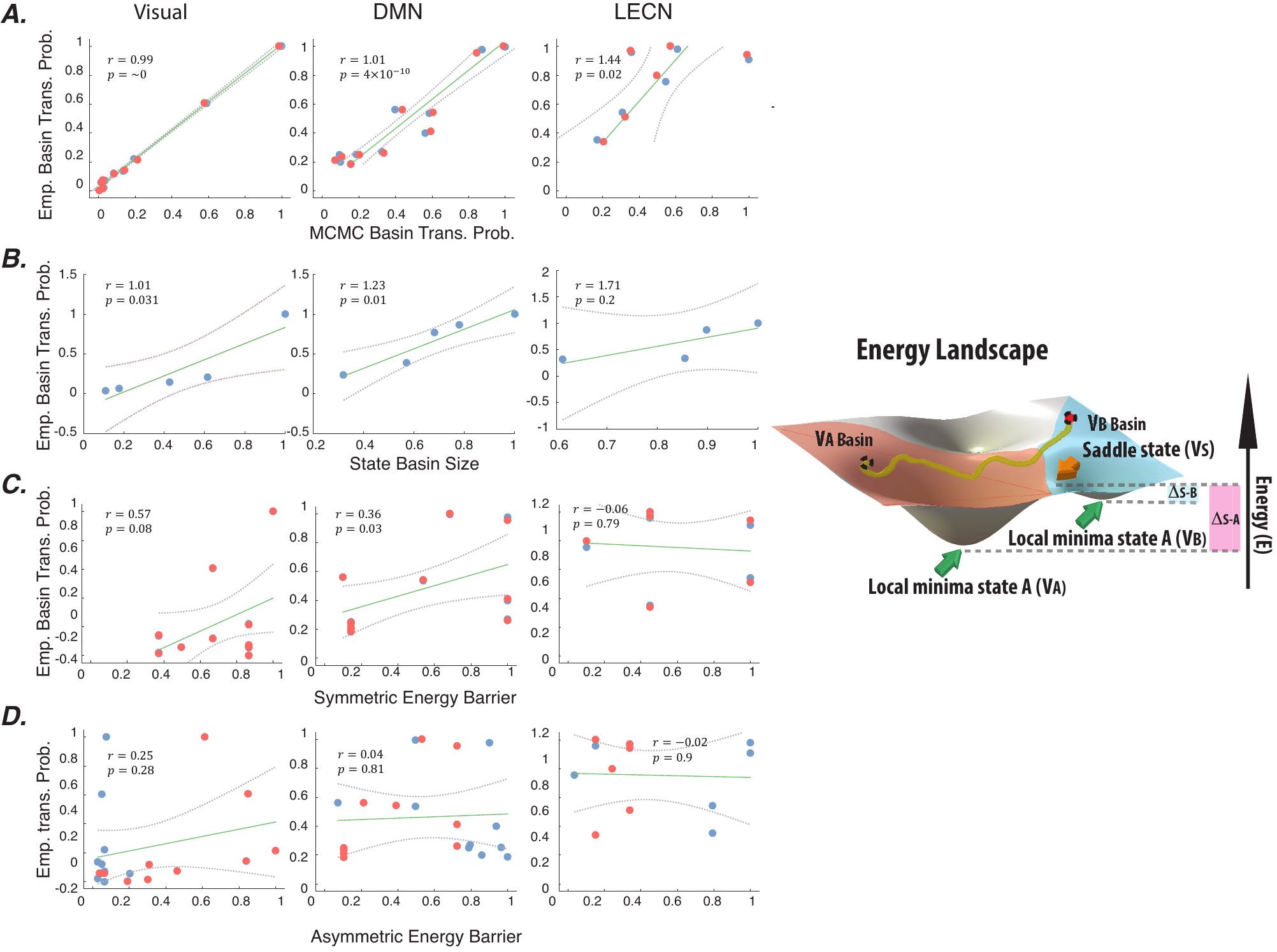}
\caption{\textbf{Empirical versus predicted probabilities of transitioning between basins.}  \emph{(A)} We used a Markov Chain Monte Carlo (MCMC) simulation over the energy landscape of each ROI via the Metropolis-Hastings algorithm to estimate theoretically predicted probabilities of transitioning between basins of local community allegiance dynamics of ROIs. The normalized empirical and model transition probabilities were significantly correlated for Class-I and Class-II ROIs but not for Class-III ROIs. Colored dots represent values on the lower (red) and upper (blue) triangles of the transition matrices. (See SI2 for details). \emph{(B)} We define the net empirical basin transition probabilities of minima states as the total in and out transition probabilities. These net probabilities are strongly correlated with the size of the basin surrounding each minima state, particularly for Class-I and Class-II ROIs. \emph{(C)} Relationship between the empirical basin transition probability and the predicted energy barriers estimated from a symmetrized transition probability matrix. \emph{(D)} Relationship between the empirical basin transition probability and the predicted energy barriers estimated from the complete asymmetric transition probability matrix. 
} 
\label{fig:Figure5}
\end{figure*}
\end{results}

\begin{discussion}

In this work we aimed to characterize the dynamic organization of large-scale brain networks and to provide a mechanistic model of the manner in which the brain transitions between large-scale functional states. Drawing on recent work demonstrating the fundamental nature of modular organization in large-scale functional dynamics \cite{sporns2016modular}, we defined these states via local patterns of brain regions' allegiance to network communities \cite{bassett2015learning}. We utilized the maximum entropy modeling (MEM) framework to estimate the probability of occurrence of these states as well as each ROI's co-occurrence with another ROI in the same community. Our results highlight the existence of three classes of ROIs with similar functional relationships. Visual, attention, sensorimotor, and subcortical ROIs tend to form a single functional community (Class-I). The remaining ROIs form the putative executive control network (Class-II) and the putative default mode and salience network (Class-III). In addition to identifying these distinct classes of ROIs that display inherently different dynamics within putative functional modules, we also studied the probabilities with which the brain transitioned from one pattern of functional modules to another pattern of functional modules. By modeling basin transitions using an MCMC random walk, we predicted empirical probabilities of state transitions with high fidelity for Class-I and Class-III ROIs, and with lower fidelity for executive control (Class-II) ROIs. Interestingly, executive control ROIs also displayed higher entropy energy landscapes, linking diverse state classes, and utilizing uniform transition probabilities across basins, consistent with their unique role in diversifying the brain's dynamic functional repertoire.  More generally, the relatively good fit of the MEM suggests that the complex patterns of network module dynamics can be described simply by pairwise interactions between regional allegiances to communities. In contrast to the widespread interest in network statistics, our results provide a the critical first steps towards a mechanistic model of brain networks dynamics.

\paragraph{Model-Based \emph{versus} Data-Driven Approaches to Studying Dynamic Functional Connectivity}

One of our fundamental aims in performing this work was to understand the dynamic functional interactions between large-scale brain networks driven by underlying neurophysiological processes at smaller spatial scales \cite{logothetis2008we}. While many empirical and data-driven approaches are currently being developed and utilized \cite{hutchison2013dynamic}, model-based approaches comprise a relatively smaller literature largely including efforts in the Virtual Brain \cite{roy2014} and dynamic causal modeling \cite{stephan2012} communities. Development of such approaches is imperative to our understanding of the role that network dynamics play in attention \cite{kucyi2016}, learning \cite{bassett2015}, language \cite{doron2012}, and memory \cite{braun2015}, and their evolution through development \cite{chai2016} or alteration in psychiatric disease \cite{seibenhuhner2013,weiss2011,du2016interaction,yu2015assessing} and neurological disorders \cite{khambhati2015,burns2014network}. Here we build on a maximum entropy based modeling framework that has been previously utilized in the context of functional activation profiles (rather than connectivity profiles) \cite{watanabe2013pairwise,watanabe2014network,watanabe2014energy}.  We adapt this approach to study the functional energy landscape of network \textit{states} and their associated attractor states dynamics. Functional brain states are defined based on regional allegiance to dynamic functional communities, providing insight into the physiological patterns of synchronization between groups of brain regions. This model-based approach revealed notable reductions in the distance $(\approx 70 \%)$ between the estimated and empirical distributions of patterns of ICN functional modules from resting state data by considering interactions between all pairs of regions. Therefore, the pair-wise MEM suggests that the observed patterns of BOLD-derived ICNs dynamic functional communities are partly due to their intrinsic tendency to synchronize with one another, likely across known structural connections \cite{watanabe2013pairwise}.

\paragraph{Attractor Communities in Dynamic Brain Networks}

Our modeling framework is explicitly based on the role that single regions play in the meso-scale organization of dynamic communities. We fit a maximum entropy model to each ROI's time series of co-occurrence with another ROI in the same community. This approach enables us to identify the dynamic functional roles that different ROIs play in network state dynamics. When combining information from all ROIs, this approach identifies network states that form local energy minima or in essence single community attractors. Indeed, while each region displays a distinct profile of activation across the energy minima, several groups of regions also show similarities in their activation profiles. We refer to these groups as classes, and observe that each single community attractor tended to contain all members of one or two ROI classes: Class-II ROIs (largely comprising executive control regions) coupled with Class-I or Class-III ROIs, whereas Class-I and Class-III ROIs coupled infrequently. This observation suggests that regions in the executive control network play unique roles in meso-scale functional dynamics \cite{andrews2014default}, forming transient control hubs that can guide interactions between large-scale functional networks \cite{mattar2015,bassett2015,braun2015,cole2013}.  

\paragraph{Dimensionality of Mesoscale Brain Dynamics}

Any assessment of brain states is faced with the question of ``So, how many are there?'' Most prior literature suggests that brain dynamics can be distilled into between 4 and 7 states at the coarsest level of inquiry \cite{allen2012,britz2010,khanna2015,shirer2012}. Yet, evidence from electrophysiology points to the presence of so-called \emph{microstates}, of which there may be many more and which can last for very short periods of time \cite{khanna2015,vakorin2011}. The repertoire of states available to the brain is therefore arguably more accurately characterized as a hierarchy, with a few coarse states composed of multiple levels of more transient (temporally localized) and focal (spatially localized) states. This complex organization requires computational and data-science approaches \cite{turkbrowne2013} such as the one that we develop in this work. Here, we uncover 25 states, as defined by local minima in the energy landscape of module dynamics during the resting state. Each state is characterized by a distinct pattern of module allegiance embodied by different brain regions. The identification of these states offers a complementary view to that provided by prior work, which has described changes in module allegiance of brain regions over time \cite{bassett2011,bassett2013,bassett2015,braun2015,mattar2015} and speculated on the cognitive drivers of these changes as regions critical for domain-general processing \cite{fedorenko2014}, associative processing \cite{bassett2013}, cognitive control \cite{bassett2015}, and cognitive flexibility \cite{braun2015}. While regional roles are important, the states themselves may also offer insights into what cognitive processes are occurring, either in parallel or in series \cite{mattar2015}. It would be interesting in future work to manipulate cognitive processes via task performance and determine the direct relationship between local minima states and mental states \cite{andrews2014default,kucyi2016,betzel2016,gu2016}. 

\paragraph{Cognitive Control, Flexibility, and Task-Switching}

Executive control networks in fronto-parietal cortices play a unique role in the pattern of results that we uncover here. Executive ROIs display states of module allegiance that are unlike the states displayed by other brain areas. Moreover, our results suggest that the brain transitions in and out of these states in a manner that is not as well-fit by an MCMC random walk on the observed energy landscape, suggesting a peculiar complexity of dynamics. Interestingly, executive ROIs also display higher entropy energy landscapes, link diverse state classes, and utilize uniform transition probabilities across basins. These findings are particularly interesting when viewed in the context of executive control function, and its instantiation in brain network architecture. Executive function supports the ability to link information to solve problems, inhibit inappropriate behaviors, and transition between tasks and states \cite{Royall2002}.  Recent work suggests that these capabilities occur via dynamic interactions between large-scale neural circuits \cite{Dajani2015,cole2013}, often taking the form of competitive or cooperative dynamics \cite{Cocchi2013} between putative functional modules \cite{mattar2015}. Indeed, recent evidence points to fronto-parietal cortices as hubs of flexible modular reconfiguration during task states \cite{bassett2013}, that are directly correlated with individual differences in learning \cite{bassett2011,bassett2015}, memory \cite{braun2015}, and cognitive flexibility \cite{braun2015}. Our current results complement these findings by suggesting that the baseline functional architecture of executive regions supports their role during task performance: (i) the high entropy energy landscapes of these regions can support highly transiently dynamics, and (ii) the uniform transition probabilities can support the integration (cooperative) and segregation (competitive) of many other cognitive systems. It is interesting to speculate that these unique features of executive region dynamics observed at rest may in part be driven by white matter microstructure, consistent with recent evidence pointing towards a structural driver of executive function \cite{gu2015}.

 \paragraph{Methodological Considerations}

For computational reasons, we have reduced the dimensionality of the data in two ways. First, we chose a small number of ROIs $(n=10)$ whose time series offer a reasonable representation of the resting state networks from which they derive (see SI). However, because of this down-sampling procedure, we are not sensitive to dynamic fluctuations \emph{within} a network. Future work using larger datasets could explore whether higher order ICA decompositions offer additional insights into finer-scale dynamics underlying the functional hierarchy that we observe here.  Second, we fit the MEM to each ROI's community allegiance time series, rather than to all ROI pairs' allegiance at the same time. Future work could aim to develop novel optimization-based methods to characterize the energy landscape of the global structure.

A distinct set of important methodological considerations relates to the dynamic community structure that we estimate as the input to the regional MEMs. We estimate dynamic community structure using a modularity maximization approach \cite{lancichinetti2011} which has an implicit structural resolution parameter that can be used to tune the number of communities. Following prior work, we employ the default parameter value of unity \cite{bassett2013robust}, and it would be interesting in future to study the changes in energy landscapes that occur at different spatial and temporal \cite{mucha2010} scales. Moreover, following the extraction of dynamic community structure, we binarize the pair-wise allegiance probabilities, necessarily losing sensitivity to fine-scale network perturbations of nodes that are loosely associated with a single community. However, since we mainly focus on the most robust attractor communities, the computational benefits of the reduced state space following binarization outweighs the cost associated with this lack of sensitivity.

See SI for complementary analyses, results, and discussions that were omitted from the manuscript for brevity. Details of the goodness of fit of the MEM, additional information on asymmetries of basin transition probabilities and basin dwell time distributions, spatial maps of ICN components, calculation of band-passed wavelet-coherence, and alternative parcellation methods are discussed in depth in SI 1,2\&6, 3, 4, and 5 respectively. 

\paragraph{Conclusions and Future Directions}

Here we present a viable mechanistic model of dynamic reconfiguration in functional brain networks estimated from resting state fMRI. By representing a brain state as a pattern of functional interactions between brain regions, we reveal structured transitions between a finite number of brain states that act as basins of attraction. Critically, each basin is characterized by a specific set of functional modules: groups of brain areas that display coherent BOLD time series. By characterizing the energy landscape surrounding these basins, we accurately predict the manner in which the brain transitions between states, and we uncover novel markers of the functional role that executive regions play in guiding these transitions. These efforts lay the groundwork for empirical investigations into how these energy landscapes change during task performance, over normative neurodeveloping, throughout healthy aging, or in the context of psychiatric disease or neurological disorders. Moreover, they lay important theoretical groundwork in the critical development of mechanistic models of brain network dynamics subserving cognitive function.

\end{discussion}

\paragraph{Acknowledgements}  This work was directly supported by the Army Research Laboratory through contract number W911NF-10-2-0022, a collaborative mechanism between J.M.V and D.S.B. In addition, D.S.B., S.G., M.G.M, and A.A. would also like to acknowledge support from the John D. and
Catherine T. MacArthur Foundation, the Alfred P. Sloan Foundation, the Army Research Office through contract number W911NF-14-1-0679, the National Institute
of Health (2-R01-DC-009209-11, 1R01HD086888-01, R01-MH107235,
R01-MH107703, and R21-M MH-106799), the Office of Naval Research, and
the National Science Foundation (BCS-1441502, CAREER PHY-1554488, and
BCS-1631550). The content is solely the responsibility of the authors
and does not necessarily represent the official views of any of the
funding agencies.


\bibliography{pnas-sample}

\end{document}